\documentclass[reprint,amsmath,amssymb,aps,]{revtex4-2}

\usepackage{tikz}
\usepackage{quantikz}
\usepackage{graphicx}
\usepackage{dcolumn}
\usepackage{bm}
\usepackage{bbold}
\usepackage{hyperref}
\usepackage{algorithm}
\usepackage{algorithmic}
\usepackage{mathtools}
\usepackage[T1]{fontenc}
\usepackage[caption=false]{subfig}
\usepackage{todonotes}
\begin{document}

\preprint{APS/123-QED}
\title{\texorpdfstring{Quantum Monte Carlo methods for Newsvendor problem \\ with Multiple Unreliable Suppliers }{My Title: Subtitle}}

\author{Monit Sharma$^{1}$}
\author{Hoong Chuin Lau$^{1,2}$}
\email{Corresponding author email: hclau@smu.edu.sg}
\address{$^1$School of Computing and Information Systems,  Singapore Management University, Singapore}
\address{$^2$Institute of High Performance Computing, A*STAR, Singapore}

\begin{abstract}
In the post-pandemic world, manufacturing enterprises face increasing uncertainties, especially with vulnerabilities in global supply chains. 
Although supply chain management has been extensively studied, the critical influence of decision-makers (DMs) in these systems remains underexplored. 
This study studies the inventory management problem under risk using the newsvendor model by incorporating DMs' risk preferences. By employing the Quantum Monte Carlo (QMC) combined with Quantum Amplitude Estimation (QAE) algorithm, the estimation of probabilities or expectation values can be done more efficiently. This offers near-quadratic speedup compared to classical Monte Carlo methods. Our findings illuminate the intricate relationship between risk-aware decision-making and inventory management, providing essential insights for enhancing supply chain resilience and adaptability in uncertain conditions.
\end{abstract}
\maketitle

\section{\label{sec:level1}Introduction}
In post-pandemic business environment, navigating uncertainties has become a critical challenge for enterprises, presenting both elevated risks and opportunities for growth. Businesses rarely operate in isolation; instead, they are embedded within complex supply chains that extend from raw material suppliers to end consumers. In a centralized supply chain, a single entity oversees and coordinates all stages, from production to distribution. Conversely, a decentralized supply chain consists of multiple independent entities, each operating autonomously and driven by self-interest, without centralized control or coordination. The distinction between centralized and decentralized supply chains has profound implications, influencing both strategic decisions and operational practices.

In decentralized supply chains, firms depend on a network of partners to meet diverse operational demands, exposing them to a wide range of uncertainties. These uncertainties can stem from typical demand fluctuations to unexpected disruptions, such as natural disasters or supply chain failures. A notable example is the reliance of Nokia and Ericsson on Philips for critical components, which highlights the vulnerability of single-sourcing strategies during unforeseen events. The lightning-induced fire at a Philips plant in New Mexico caused significant production disruptions, leading to substantial financial losses for Ericsson and ultimately contributing to its exit from the cellphone market. In contrast, Nokia's agile response, quickly shifting to alternate suppliers, demonstrates the importance of resilience in mitigating supply chain risks. These cases emphasize the critical role of proactive risk management strategies in safeguarding supply chain operations and ensuring business continuity~\cite{chopra}.

Effectively managing risk is increasingly recognized as a critical factor for achieving a competitive advantage. Supply chain risks can arise from various sources, each distinct to its specific context. Although it would be ideal to address all these risks simultaneously, doing so is fraught with challenges. Even managing risks within a single category presents its own complexities and has received varying levels of attention in the research literature \cite{khan2007risk, peck2006reconciling, ho2015supply}. This paper focuses on inventory management, a vital component of supply chain management. Inventory management primarily addresses two key questions: when to order and how much to order. This study concentrates specifically on the latter.

Implementing an incorrect inventory policy can result in unfavorable outcomes, such as excess or insufficient inventory, leading to increased costs and diminished profits. It is crucial to note that these policies are crafted by management, who base their decisions on an evaluation of factors including market conditions, competitive dynamics, costs, and prices. Essentially, management's decisions are informed by their perception of the opportunities and risks present in an uncertain operational environment.

Stochastic optimization \cite{SBO}, addresses inherent uncertainty by integrating probabilistic elements into the optimization process. This approach has gained increasing importance, particularly in managing complex systems influenced by uncontrollable variables such as weather or major events like wars and pandemics. Traditional optimization methods often struggle to manage the complexities introduced by these probabilistic factors.

In this context, Monte Carlo methods \cite{monte}, are widely recognized as a preferred approach, employing statistical techniques and random sampling to approximate solutions for complex problems. Particularly effective in addressing stochastic optimization challenges, Monte Carlo methods excel where deterministic models falter due to the unpredictable nature of such problems. By generating random samples from relevant distributions, these techniques introduce randomness into the optimization process, facilitating the exploration of a broad spectrum of potential outcomes.

Monte Carlo simulations often present significant and demanding computational challenges, particularly when dealing with complex systems or optimization problems in high-dimensional spaces. Quantum computing, which leverages the principles of quantum mechanics, offers a transformative approach to complex information processing tasks, representing a significant paradigm shift in problem-solving methodologies \cite{qcqi}. It holds the potential for algorithmic speedup across a range of tasks, including factorization \cite{Shor_1997} and optimization \cite{grover,funopt,QAE,markov}. 

Grover's search algorithm \cite{grover}, theoretically provides a quadratic speedup for searching unstructured databases. While a classical computer requires $\mathcal{O}(N)$ computational steps to find a solution in a database of size $N$ with high probability, a quantum computer accomplishes this in $\mathcal{O}(\sqrt{N})$ steps. Grover's algorithm has been extended and adapted for function optimization \cite{funopt}, amplitude amplification and estimation \cite{QAE}, and Markov chain algorithms \cite{markov}. Notably, the amplitude estimation algorithm offers near-quadratic speedups for estimating expectation values, potentially outperforming classical Monte Carlo methods in specific scenarios. And by employing Quantum Monte Carlo (QMC) combined with Quantum Amplitude Estimation (QAE) algorithm, the estimation of probabilities or expectation values can be done more efficiently, which offers near-quadratic speedup \cite{Montanaro_2015} compared to classical Monte Carlo methods.

This study seeks to bridge existing gaps in inventory management under uncertainty by introducing a decision-maker with heightened risk awareness. Building upon the traditional newsvendor model, we investigate the effects of integrating a decision-maker with specific risk preferences. To support this analysis, we employ QAE \cite{QAE} as a critical tool, providing insights into the complex dynamics of decision-making in stochastic inventory settings. This work naturally extends from \cite{sharma2024quantumenhancedsimulationbasedoptimizationnewsvendor}, where we examined a 100\% reliable newsvendor, to a scenario involving multiple unreliable newsvendors, where the variability in reliability introduces risk.

In Section \ref{sec:level2}, we provide a comprehensive overview of the Newsvendor problem, examining the implications of both unreliable and perfectly reliable suppliers. We also detail the formulation of the problem as a profit function for $N$ suppliers, each defined by unique costs and reliability parameters. Moving to Section \ref{sec:level3}, we explore QAE in depth, emphasizing its advantages over classical Monte Carlo simulations. Section \ref{sec:level4} focuses on our application of Quantum Monte Carlo (QMC) techniques. In Section \ref{sec:level5} we construct the quantum circuit and highlight the steps involved in the computation. In Section \ref{sec:level7}, we present the results of our experiments across various random demand and real world scenarios. Finally, Section \ref{sec:level8} provides insights into future research directions.

\section{\label{sec:level2} Newsvendor Problem with unreliable suppliers}

We analyzed the scenario of a newsvendor ordering items from multiple suppliers, categorized as either perfectly reliable or unreliable. A perfectly reliable supplier consistently delivers the exact amount requested, as is typical in the standard newsvendor problem \cite{sharma2024quantumenhancedsimulationbasedoptimizationnewsvendor}. In contrast, an unreliable supplier has a probability of delivering less than the requested quantity.

By definition, a newsvendor faces the crucial decision of determining the quantity of product to procure from its supplier for a single selling season, without prior knowledge of the random demand \cite{edgeworth1888mathematical}. Consequently, the newsvendor's sales throughout the season are influenced by both the actual demand that materializes and the quantity of stock supplied. This situation leads to two distinct economic outcomes: If actual demand exceeds supply, the newsvendor sells out its entire inventory but incurs unmet demand. Conversely, if supply exceeds demand, the newsvendor meets all demand but faces excess stock. The conventional critical fractile solution suggests that to optimize expected profit over the selling season, the newsvendor should set the supply level so that the probability of meeting demand equals the ratio of the marginal cost of understocking to the sum of the marginal costs of overstocking and understocking \cite{stevenson2014operations}. See Appendix \ref{form} for further details.

Initially, the newsvendor model assumes a single supplier with a deterministic capacity. This foundational framework extends in two significant ways:

\begin{enumerate}
    \item \textbf{Multiple Suppliers with Deterministic Capacities}: The newsvendor should rank suppliers based on cost. In the absence of administrative costs, the newsvendor should start ordering from the least expensive supplier. If this supplier’s capacity is insufficient, the process continues with the next least expensive supplier until an adequate order is placed.

    \item \textbf{Single Supplier with Uncertain Capacity}: If the newsvendor pays only for the quantity received, it should order at least as much as it would if the supplier's capacity were deterministic. However, if the supplier's random capacity is influenced by the size of the order \cite{gerchack}, the newsvendor may need to adjust its order upwards to account for potential capacity variability.
\end{enumerate}

Given these two natural extensions of the basic newsvendor model, the central question of this paper arises: What are the implications when these models are combined? To what extent do the key insights from each extension remain applicable? For example, in the multiple supplier case, the newsvendor typically avoids ordering from a supplier if the same unit can be obtained at a lower cost from another source. However, how does this apply when the newsvendor is uncertain about the alternative sources offering lower costs? 

How should the newsvendor decide which suppliers to place orders with and which to avoid? Similarly, in the case of the uncertain capacity variant, the model guides the ordering quantity when no alternative sources are available. But if there are alternative, potentially unreliable sources, how does the newsvendor determine the order quantity from each supplier?

This paper aims to address these questions by examining whether orders should be placed with specific suppliers and, if so, in what quantities, within a multi-supplier framework that incorporates procurement and reliability considerations \cite{dadam, feng}.

\subsection{Problem Notation}
The notation that is employed in describing the model is introduced below:

\begin{align*}
    N & :\ \ \text{Number of Suppliers} \\
    x_i & :\ \  \text{Binary decision variable, } x_i = 1 \\ 
     &\ \ \ \ \ \text{if ordered from supplier } i , \text{else } x_i = 0 \\
     q_i & : \ \ \text{Decision variable for order quantity from supplier } i\\
     R_i & : \ \ \text{Random variable denoting the reliability of supplier } i\\
     c_i & : \ \ \text{Unit ordering cost from supplier } i \\
     F_i & : \ \ \text{Fixed ordering cost from supplier } i \\
     K_i & : \ \ \text{Supply capacity of supplier } i\\
     D & : \ \ \text{Random demand to be satisfied by the newsvendor}\\
     p & : \ \ \text{price to be charged for each unit}\\
     o & : \ \ \text{unit shortage cost, should always be } o > c_i\\
     w & : \ \ \text{unit salvage value for unsold product,} \\ 
     &\ \ \  \text{ should always follow } c_i > w
\end{align*}

\subsection{Problem Formulation}
We let $f(q,RO,D)$ be the profit function for deciding $(q_1, q_2, ..., q_N)$ and $(x_1, x_2, ...,x_N)$ with the total received order $RO = \sum_{i=1}^{N} R_i q_i$ and demand $D$,\\
if $RO>D$:
\begin{equation}\label{eq:one}
    f(q,RO,D) = pD - \sum_{i=1}^{N} c_i R_i q_i - \sum_{i=1}^{N} F_i x_i + w(RO-D)
\end{equation}

else if $D \ge RO$
\begin{equation}\label{eq:two}
    f(q,RO,D) = pD - \sum_{i=1}^{N} c_i R_i q_i - \sum_{i=1}^{N} F_i x_i - o(D-RO)
\end{equation}

which can be combined into one profit function:

\begin{align}
    f(q,RO,D) = & (p-w)D + \sum_{i=1}^{N} (w-c_i)R_i q_i \\\label{eq:function} 
    & - (o-w) \Bigg[\Bigg(D- \sum_{i=1}^{N} R_i q_i \Bigg)^{+} \Bigg] - \sum_{i=1}^{N} F_i x_i \notag 
\end{align}

We define the newsvendor's decision problem as follows:
Before the start of a single selling season, the newsvendor must select a supplier from a pool of $N$ independent suppliers. Each supplier offers inventory at a different cost, denoted by $c_i$, representing the per unit purchase cost. The newsvendor's total available stock for the season, used to fulfill demand ($D$), is the sum of inventory delivered by all $N$ suppliers ($\sum_{i=1}^{N} R_i q_i $). The newsvendor then sells as much of this stock as demand allows $(\text{min }\{D, RO\})$, at the per unit selling price $p$. Any excess stock above realized demand $(RO-D)$ is salvaged at a per unit value $w<c_i$, while any shortages $(D-RO)$ are assessed a per unit penalty cost $o>c_i$.

The objective of the newsvendor is to determine a vector of non-negative order quantities $\textbf{q} = {q_1,q_2,...,q_N}$ to maximize $f(q, RO, D)$, representing its expected profit for the selling season. This profit is calculated as the sum of expected sales and salvage revenues, subtracting the expected shortage and purchase costs.

\section{\label{sec:level3}Quantum Amplitude Estimation (QAE)}

QAE was first introduced in \cite{QAE}, and is a quantum algorithm that gives a quadratic speedup compared to Monte Carlo simulations traditionally used on classical computers. It assumes that the problem of interest is given by a unitary operator $\mathcal{A}$ acting on $n+1$ qubits such that

\begin{equation}\label{eq:4}
    \mathcal{A}|0\rangle_n |0\rangle= \sqrt{1-a}||\Psi_0\rangle_n |0\rangle + \sqrt{a}|\Psi_1\rangle_n |1\rangle
\end{equation}

where $a \in [0,1]$ and $|\Psi_0\rangle$ and $|\Psi_1\rangle$ are two normalized and orthogonal states.

QAE enables the estimation of the parameter $a$ with a high probability, resulting in an estimation error that scales as $\mathcal{O}(1/M)$, where $M$ represents the number of $\mathcal{A}$ operator applications. The key component in this process is the Grover operator $\mathcal{Q}$, which is constructed as follows:

\begin{equation*}
    \mathcal{Q} = \mathcal{A} S_0 \mathcal{A^{\dagger}} S_{\Psi_0}
\end{equation*}
Here, $S_{\Psi_0} = \mathbb{I} - 2 |\Psi_0 \rangle \langle \Psi_0| \otimes |0\rangle \langle 0|$ and $S_0 = \mathbb{I} - 2|0\rangle_{n+1} \langle 0|_{n+1}$, as elaborated in detail in \cite{QAE}. The applications of $\mathcal{Q}$ are commonly referred to as \textit{oracle queries}.

\begin{figure}
    \centering
    \includegraphics[width = \linewidth]{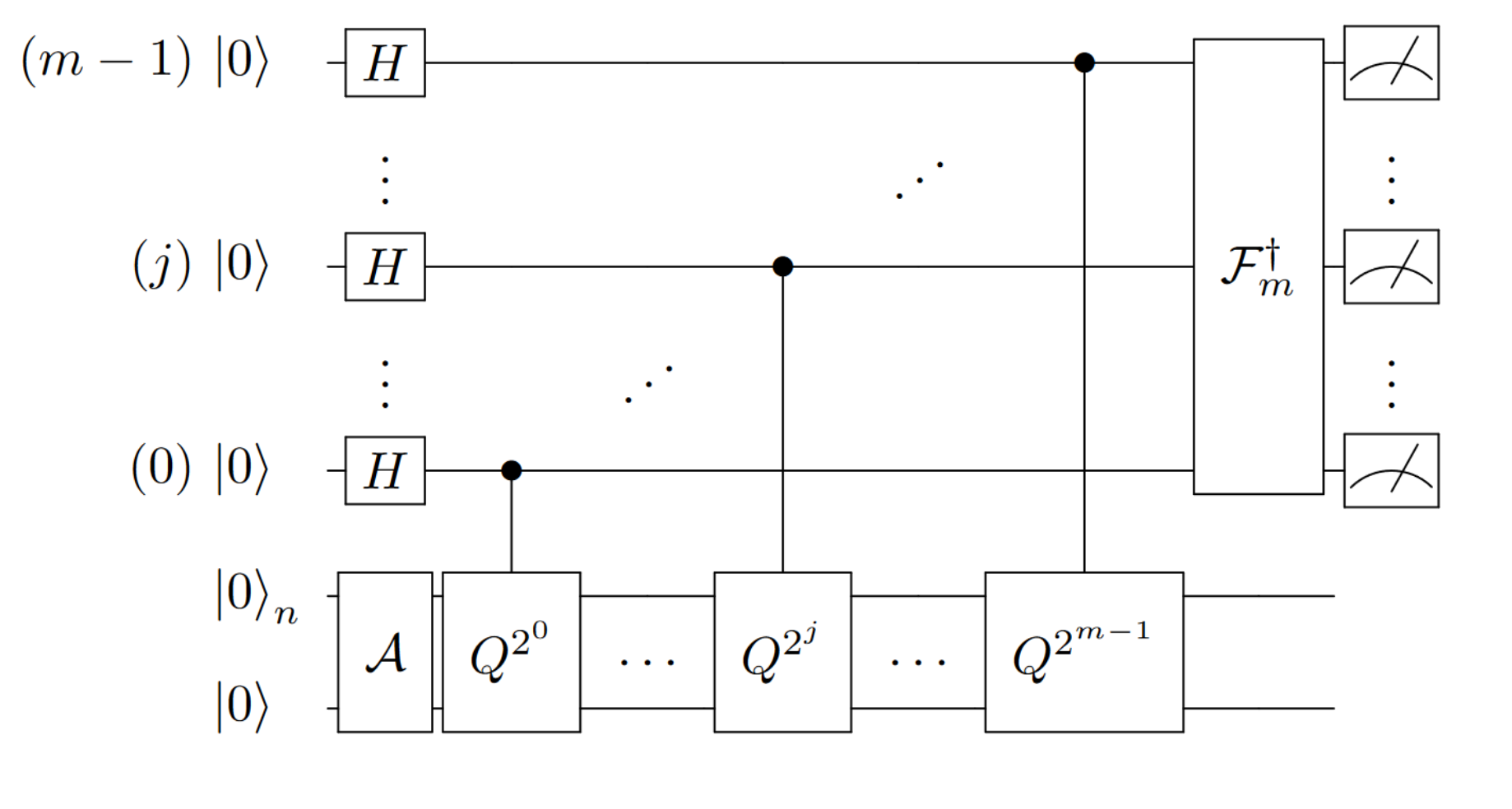}
    \caption{QAE circuit with $m$ ancilla qubits and $n+1$ state qubits. $H$ is the Hadamard gate and $\mathcal{F}^{\dagger}_m$ denotes the inverse Quantum Fourier Transform on $m$ qubits.}
    \label{fig:1}
\end{figure}

The standard form of QAE is derived from Quantum Phase Estimation (QPE) \cite{qpe}. It incorporates an additional set of $m$ ancillary qubits, all initially set in an equal superposition state. These ancillary qubits play a crucial role in representing the outcome. The number of quantum samples is defined as $M = 2^m$, and the algorithm involves the application of geometrically increasing powers of the $\mathcal{Q}$ operator, controlled by the ancillary qubits. Finally, it culminates in executing a Quantum Fourier Transform (QFT) on the ancillary qubits just before measurement, as depicted in Figure \ref{fig:1}.

Following this, the integer measurement result, denoted as $y$ and falling within the range ${0, \ldots, M-1}$, is converted into an angle $\tilde{\theta}_a$ using the formula $\tilde{\theta}_a = \frac{y \pi}{M}$. Subsequently, the estimated value of $a$, denoted as $\tilde{a}$, is defined as $\tilde{a} = \sin^2 (\tilde{\theta}_a)$, which lies within the interval $[0, 1]$. The estimator $\tilde{a}$ satisfies the following relationship:

\begin{align}\label{eq:5}
    |a -\tilde{a}| &\le \frac{2\sqrt{a(1-a) \pi}}{M} + \frac{\pi^2}{M^2} \\
    & \le \frac{\pi}{M} + \frac{\pi^2}{M^2} = \mathcal{O}(M^{-1})
\end{align}

with probability of at least $\frac{8}{\pi^2} \approx 81\%$. This demonstrates a quadratic speedup when compared to the $\mathcal{O}(M^{-1/2})$ convergence rate of classical Monte Carlo methods \cite{QAE}. The probability of success can be significantly enhanced, approaching close to $100\%$, by repeating this process multiple times and utilizing the median estimate \cite{risk}. These estimates $\tilde{a}$ are confined to the grid ${\sin^2 (y \pi/M) : y = 0,....,M/2}$ based on the potential measurement outcomes of $y$. Due to the symmetry of the sine function, the algorithm's output $\tilde{a}$ resides on a grid with $M/2 + 1$ possible values within the range of $[0,1]$.

The standard Quantum Amplitude Estimation (QAE) method often leads to complex quantum circuits and provides discrete estimates, represented as $\tilde{a}$, depending on the evaluation qubits $m$. Consequently, recent progress has introduced various QAE variations aiming to improve both the accuracy and complexity of the algorithm \cite{simple,raymond, counting, iter, faster}. These alternative methods offer a continuous range of estimated values while simplifying the circuit by removing the necessity for ancilla qubits and the Quantum Fourier Transform (QFT). Significantly, due to their ability to generate estimates over a continuous spectrum, these QAE variations are better suited for Simulation-Based Optimization (SBO) tasks.

All variants of QAE, that do not make use of QPE are based on the fact that

\begin{equation}\label{eq:7}
\begin{split}
    \mathcal{Q}^k \mathcal{A}|0\rangle_n |0\rangle =& \cos((2k+1)\theta_a |\Psi_0\rangle_n |0\rangle  + \\
    & \sin((2k+1)\theta_a)|\Psi_1\rangle_n |1\rangle
\end{split}
\end{equation}

where $\theta_a$ is defined as $a = \sin^2(\theta_a)$. In other words, the probability of measuring $|1\rangle$ in the last qubit is given by 

\begin{equation*}
    \mathbb{P}[|1\rangle] = \sin^2 ((2k+1) \theta_a)
\end{equation*}

The algorithms primarily vary in their approaches to determining the various values for the powers $k$ of $\mathcal{Q}$ and how they aggregate the results into the final estimate of $a$. We will employ the Iterative Quantum Amplitude Estimation (IQAE) \cite{iter} algorithm.

QAE is used to estimate an expected value $\mathbb{E}[f(D)]$, for a given random variable $D$ and a function $f: \mathbb{R} \rightarrow [0,1]$. Expected values in this form commonly appear as objective functions in Simulation-Based Optimizations (SBO).

\
Suppose a quantum operator $\mathcal{P}_X$ that acts like 
\begin{equation}\label{eq:8}
    \mathcal{P}_D |0\rangle_n = |\psi\rangle_n = \sum_{i = 0}^{N-1} \sqrt{p_i} |i\rangle_n
\end{equation}
where the probability of measuring the state $|i\rangle_n$ is $p_i \in [0.1]$ with $\sum_{i=0}^{N-1} p_i = 1$ and $N = 2^n$. The state $|i\rangle_n$ is one of the $N$ possible realizations of a bounded discrete random variable $D$ which for instance represents a discretized demand for newspaper stocks. We load the discretized probabilities $p_i$ into the amplitude of $n$ qubits employing this operator $\mathcal{P}_D$.

Creating a qubit register to approximate a probability density function (PDF) can be accomplished through quantum arithmetic, provided that the function is efficiently integrable \cite{inter}. Alternatively, for all smooth and differentiable functions, this can be achieved using matrix product states \cite{smooth}. Another approach involves utilizing quantum generative adversarial networks (qGANs) to approximate generic functions \cite{qgan}.

We consider a function $f: \{0,...,N-1\} \rightarrow [0,1]$ and a corresponding operator:
\begin{equation}\label{eq:9}
    F|i\rangle_n |0\rangle = \sqrt{1-f(i)}|i\rangle_n |0\rangle + \sqrt{f(i)}|i\rangle_n |1\rangle
\end{equation}

 for all $i \in \{0,....,N-1\}$ acting on an ancilla qubit.

 An operator like $F$ can be created through the utilization of quantum arithmetic and related methods \cite{efficient, opquan}. In our research, we will adopt the approach outlined in a previous work \cite{risk,qsbo}. This approach involves approximating $f$ using a Taylor expansion and utilizing controlled Pauli rotations to map the function values onto the qubit amplitudes. This mapping allows for the flexibility to strike a balance between precision and circuit complexity. This equilibrium can be attained by selecting an appropriate number of Taylor terms to approximate the function, thus avoiding the need for intricate quantum arithmetic operations.

Next, suppose a discrete random variable $X$ taking values in $\Omega_D = \{d_i\}^{N-1}_{i=0}$, where $N= 2^n$ for a given $n$, with the corresponding probabilities $p_{d_i} = \mathbb{P}[D=d_i]$. Then the expectation value can be written as 

\begin{equation}\label{eq:10}
    \mathbb{E}[f(D)] = \sum_{i=0}^{N-1} p_{d_i} f(d_i) = \sum_{\hat{d} = 0}^{N-1} p_{\phi(\hat{d})} f(\phi(\hat{d})) 
\end{equation}

where $\phi : \{0,...,N-1\} \rightarrow \Omega_D$ represent the affine transformation from $\hat{d} \in \{0,...,N-1\}$ to $d \in \Omega_D$.

Now we can encode $\mathbb{E}[f(D)]$ in $\mathcal{A}$ making use of Eq. (\ref{eq:9}) and Eq. (\ref{eq:10}). First we load the discretized probabilities using Eq. (\ref{eq:8}) into the amplitudes of $n$ qubits by means of $\mathcal{P}_D$

Then we add one more qubit and use $F$ from Eq. (\ref{eq:9}) to define $\mathcal{A} = F(\mathcal{P}_D \otimes \mathbb{I})$. The state after applying it is given by:

\begin{equation}\label{eq:11}
\begin{split}
    \mathcal{A}|0\rangle_n |0\rangle  =&    \sum_{i=0}^{N-1} \sqrt{1-f(i)} \sqrt{p_i}|i\rangle_n|0\rangle \\
    +& \sum_{i=0}^{N-1} \sqrt{f(i)} \sqrt{p_i}|i\rangle_n|1\rangle 
\end{split}
\end{equation}

We use amplitude estimation to approximate the probability of measuring $|1\rangle$ on the last qubit, this implies that 

$$ a = \sum_{i=0}^{N-1} p_i f(i)$$
which is the desired expected value from Eq. (\ref{eq:10})

We can also make use of QAE to estimate variance, cumulative distribution functions, and the (Conditional) Value at Risk.

\section{\label{sec:level4}Quantum Monte Carlo}

 QAE is a quantum algorithm that provides a generic approach for measuring an expectation value with a quadratic improvement in efficiency over classical methods. In this section, we discuss QMC methods \cite{quantmonte}.

Consider a discrete random variable $D$ with values in $\Omega_D$, a decision variable $q$ in $\mathbb{R}^N$, and a function $f : \Omega_D \times \mathbb{R}^N \rightarrow \mathbb{R}$. Instead of employing a Monte Carlo simulation (see Appendix \ref{montecarlo} for additional information), we use the QAE algorithm to compute the expectation value.

The expectation value $\mathbb{E}[f(D,q)]$ can be evaluated by preparing the operator $\mathcal{A}$ as described in Eq. (\ref{eq:11}). Here, the value of $q$ is treated as a parameter of the function $f$, i.e., $f = f_q(D)$ and $F = F_q$.

$$F_q |d\rangle |0\rangle = \sqrt{1-f_q(d)}|d\rangle |0\rangle + \sqrt{f_q(d)}|d\rangle|1\rangle $$

\begin{figure}
    \centering
    \includegraphics[width=\linewidth]{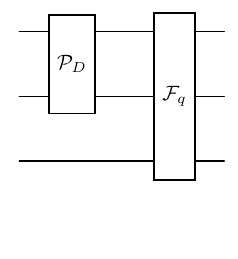}
    \caption{The circuit diagram implementing the $\mathcal{A}$ operator , the decision variable $q$ is a parameter of the function mapping $\mathcal{F}_q$. The first two-qubit contains the probability distribution, while the final qubit is the objective qubit. }
    \label{fig:a_operator}
\end{figure}

The action of $\mathcal{A} = \mathcal{A}_q$, as shown in Fig. \ref{fig:a_operator} can be formulated as 

$$\mathcal{A}_q |0\rangle^{\otimes (n+1)} = F_q ( \mathcal{P}_D \otimes \mathbb{I}_1) |0\rangle^{\otimes n} |0\rangle$$

In this setup, the upper $n$ qubits represent the random variable $D$, while the last qubit is dedicated to applying $F_q$ and distinguishing between states $|\Psi_0\rangle$ and $|\Psi_1\rangle$ by marking them with $|0\rangle$ and $|1\rangle$, respectively. This qubit, designated as the objective qubit, differentiates between good and bad states.

A single assessment of the objective function involves three main steps: initializing the parameterized operator $\mathcal{A}_q$, creating the corresponding $\mathcal{Q}_q$, and executing QAE to derive an approximation $\tilde{a}_q \approx \mathbb{E}[f(D,q)]$. The success probability can be boosted by repeating the experiment multiple times.

In our study, we employ the Iterative Quantum Amplitude Estimation (IQAE) method \cite{iter}. We benchmark its performance across various demand scenarios to evaluate its effectiveness in addressing the problem at hand.

\section{\label{sec:level5}Quantum Circuit}

To model ans solve the newsvendor problem with unreliable suppliers on a gate-based quantum computer, several fundamental components must be addressed. These include: the probability distribution that governs the evolution of random variables $D$ within the quantum circuit, the formulation of the operator used to compute the payoff function, and the calculation of the expectation value associated with this payoff.

\subsection{Distribution Loading}
The initial component of our model involves a circuit operating on a probability distribution derived from historical sales data, which is used to represent current demand. This circuit initializes the distribution onto a quantum register, where each basis state represents a potential value and its amplitude corresponds to the associated probability. Specifically, given an $n$-qubit register, demand data $\{D_i\}$ for $i \in \{0, \ldots, 2^n -1\}$ and corresponding probabilities $\{p_i\}$, the distribution loading module prepares the state:

\begin{equation*}
   |\psi\rangle_n = \sum_{i = 0}^{N-1} \sqrt{p_i} |i\rangle_n
\end{equation*}

Analytical formulas used in option pricing \cite{option} and newsvendor models \cite{qsbo} typically assume that the underlying data follows a log-normal distribution. As shown in \cite{inter}, log-concave probability distributions can be efficiently loaded onto a gate-based quantum computer. However, these simplified assumptions often fail to capture critical market dynamics, limiting the model’s applicability to real-world scenarios. Thus, accurately capturing the market-implied probability distribution is essential for valuation models to estimate intrinsic value effectively.

Loading arbitrary states into quantum systems generally requires an exponential number of gates \cite{expo}, making it impractical to model arbitrary distributions directly as quantum gates. Given that distributions of interest often have specific forms, this challenge can be addressed using quantum Generative Adversarial Networks (qGANs) \cite{qgan}. qGANs enable the efficient loading of a distribution with a polynomial number of gates. These networks can learn the random distribution $X$ underlying observed data samples $\{d^0, d^1, \ldots, d^{k-1}\}$ and directly encode it into a quantum state.

After the training, the output of the generator of a quantum state is

\begin{equation}\label{eq:12}
    |\psi(\theta_p)\rangle_n = \sum_{i=0}^{2^{n}-1} \sqrt{p_i (\theta_p)}|i\rangle_n
\end{equation}

that represents the target distribution. The $n-$qubit state $|i\rangle_n = |i_{n-1}....,i_{0} \rangle$ encodes the integer $i = 2^{n-1}i_{n-1} + ....+ 2i_1 + i_0 \in \{0,..,2^n -1\}$ with $i_k \in \{0,1\}$ and $k=0,....,n-1$. The probabilities $p_i(\theta_p)$ approximate the random distribution underlying the training data. We note that the outcomes of a random variable $D$ can be mapped to the integer set $\{0,...,2^n -1\}$ using an affine mapping. This approach can be easily extended to multivariate data, where we use a separate register of qubits for each dimension.

\subsection{Computing the Payoff}

We obtain the expectation value of a linear function $f$ of a random variable $D$ with QAE by creating the operator $\mathcal{A}$ such that $a = \mathbb{E}[f(D)]$ (see Eq.(\ref{eq:10})). Once $\mathcal{A}$ is implemented we can prepare the state in Eq.(\ref{eq:4}), and the $\mathcal{Q}$ operator, which only requires $\mathcal{A}$ and generic quantum operations.

The payoff function for the newsvendor is loaded as a conditional operation from Eq. \ref{eq:one} and \ref{eq:two} , $f: \{0,...,2^n-1\} \rightarrow[0,1]$, which we can write as $f(i) =f_1 i + f_0$. We can efficiently create an operator $F$ that performs Eq.(\ref{eq:9}), and making use of controlled $Y$-rotations give us:

\begin{equation}{\label{eq:13}}
    |i\rangle_n |0\rangle \rightarrow|i\rangle_n \big(\cos [f(i)]|0\rangle + \sin [f(i)]|1\rangle \big)
\end{equation}

\begin{figure}[h!]
    \centering
    \includegraphics[width =\linewidth]{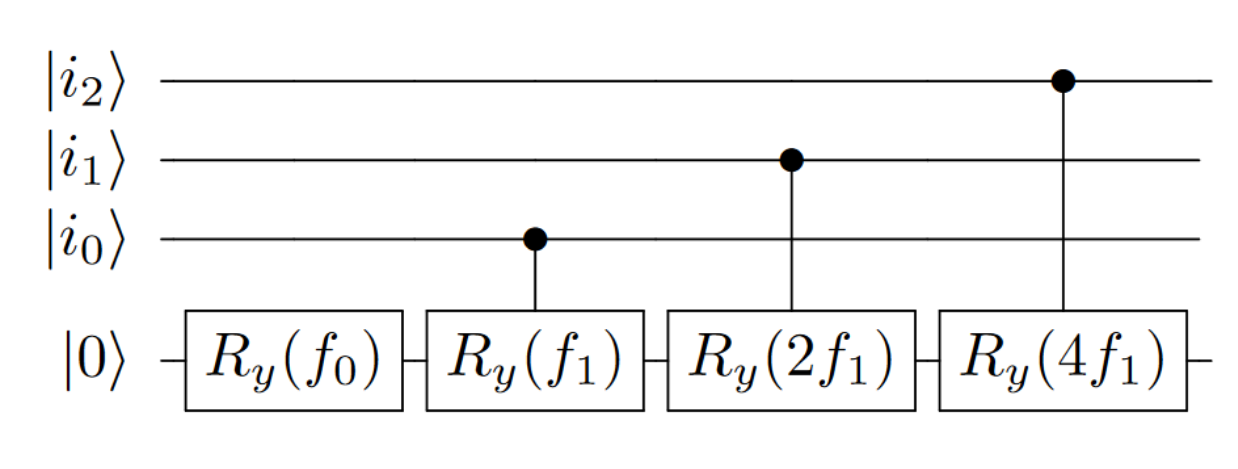}
    \caption{Quantum circuit that loads the objective value onto the final qubit, efficiently creating the state in Eq.(\ref{eq:13})}
    \label{fig:3}
\end{figure}

To implement the linear term of $f(i)$ each qubit $j$ (where $j \in \{0,...,n-1\})$ in the $|i\rangle_n$ register acts as a control for a Y-rotation with angle $2^j f_1$ of the ancilla qubit. A rotation of the ancilla qubit implements the constant term $f_0$ without any controls as shown in Fig.(\ref{fig:3}). 

The operator $\mathcal{A}$ then, acts as 

\begin{equation}
\begin{split}
    \mathcal{A}|i\rangle_n |0\rangle = \sum_{i=0}^{2^n-1} \sqrt{p_i (\theta_p)} \cos [f(i)]|i\rangle_n |0\rangle \\ 
    +  \sum_{i=0}^{2^n-1} \sqrt{p_i (\theta_p)} \sin [f(i)]|i\rangle_n |1\rangle
\end{split}
\end{equation}

and we can use it within QAE. The output of QAE is then transformed to an estimate of the expectation value by reverting the applied scaling. See Appendix \ref{payoff} for more details.

\section{\label{sec:level7}Results}

The experiment was conducted across various demand distributions while maintaining a consistent set of hyperparameters, utilizing the Qiskit framework \cite{qiskit}. These parameters included two suppliers ($N=2$), each with distinct procurement costs ($c_1 = 0.95$, $c_2 = 0.80$) and fixed expenses ($f_1 = 0.03$, $f_2 = 0.04$), while keeping selling price ($p = 1.4$), salvage value ($w = 0.6$), and overage cost ($o = 1.3$) consistent.

Initially, the reliability of both suppliers was modeled as a random variable, following a normal distribution with mean $\mu_i$ and variance $\nu_i$, where $0 \le \mu_i \mp 2 \sqrt{\nu_i} \le 1$. In subsequent tests, the variance was held constant at $0.1$, while the mean of the random variable was systematically varied from $0.1$ to $1.0$ in uniform increments. A reliability of $0.7$, for example, indicates that the supplier is $70\%$ reliable.

This model was implemented on the noise-free \textit{ibm qasm simulator}. For our specific scenario, we employed different random distributions, the underlying function of which remained unknown. This distribution was encoded onto $n = 4$ qubits and truncated within the range $\Omega_D = [0, 15]$. Since $2^n$ values can be represented using $n$ qubits, this framework can be scaled to $\Omega_D = [0, 2^n -1]$ by utilizing $n$ qubits to encode the demand value.

\subsection{Demand Scenario 1}

We generated a random demand ranging between $\{0,15\}$. Subsequently, we utilized the qGAN to learn this distribution and efficiently loaded it onto the quantum circuit.

\begin{figure}[h!]
    \centering
    \includegraphics[width = \linewidth]{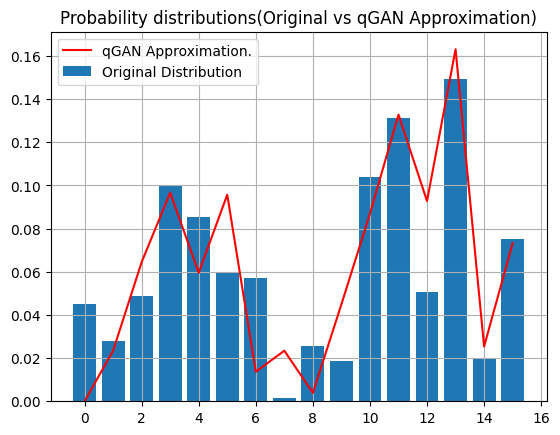}
    \caption{\textbf{Demand Scenario 1}: Comparison of the qGAN learned distribution with the randomly generated demand distribution.}
    \label{fig:demand_qgan_1}
\end{figure}
The optimization process employs COBYLA, with entropy loss as the loss function. The results illustrated in Fig. \ref{fig:demand_qgan_1} were derived from qGAN training over $200$ epochs. To enhance training robustness, the optimizer’s learning rate was set to $10^{-3}$. Given this higher learning rate, a training duration of $200$ optimization epochs is adequate.

\begin{figure}[h!]
    \centering
    \includegraphics[width=\linewidth]{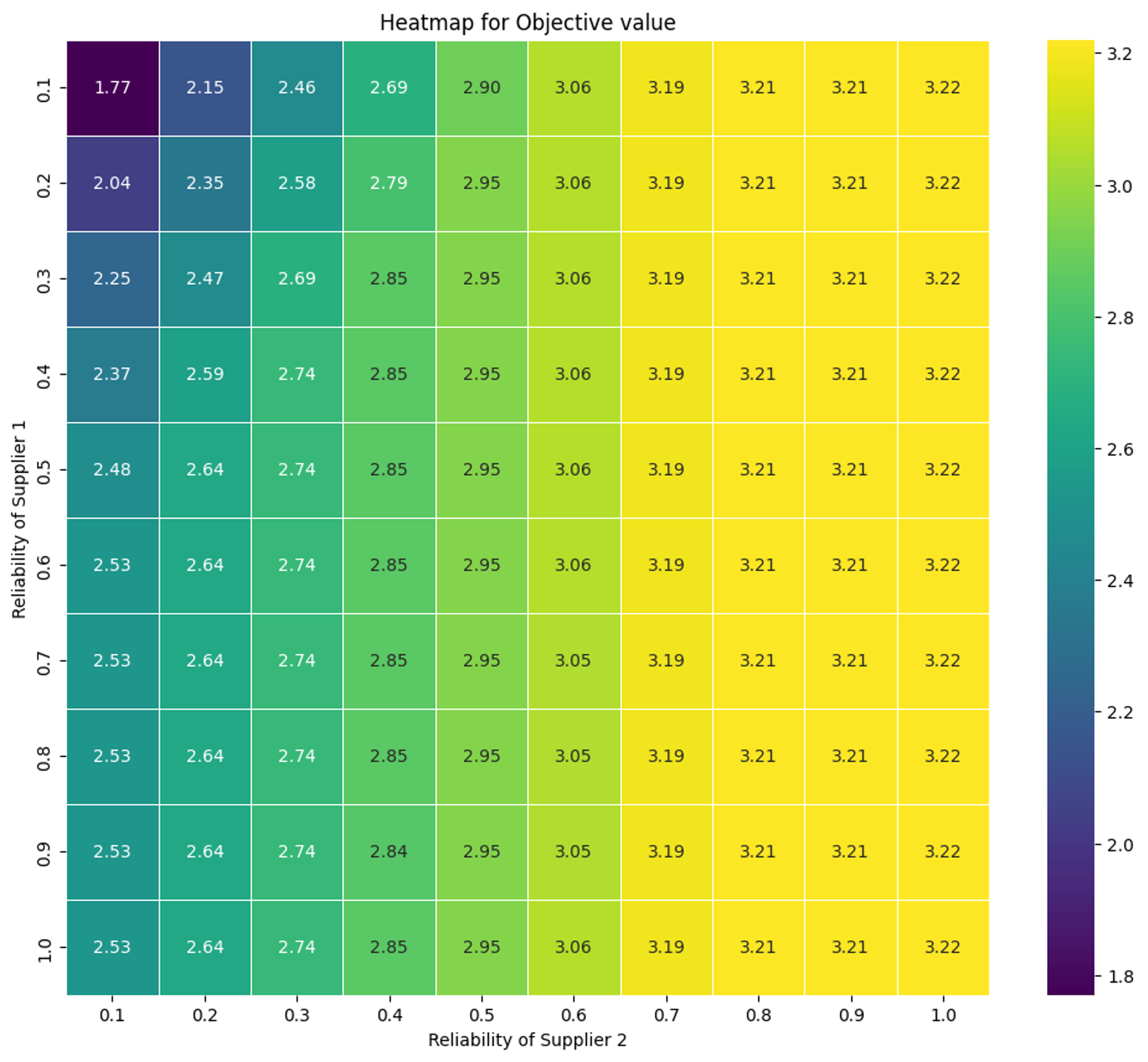}
    \caption{\textbf{Objective Value}: The realized profit for Demand Scenario 1 is shown in the heatmap. The $(x, y)$-axis represents the reliability of the suppliers, with individual cells indicating the realized profit corresponding to the specific reliability levels of each supplier.}
    \label{fig:objective1}
\end{figure}

We performed experiments by varying the reliability factor $R_i$ for both suppliers and presented the results in the form of a heatmap. This heatmap displays the objective value (profit) across different reliability factors for each supplier (see Fig. \ref{fig:objective1}).

\begin{figure}[h!]
    \centering
    \includegraphics[width=\linewidth]{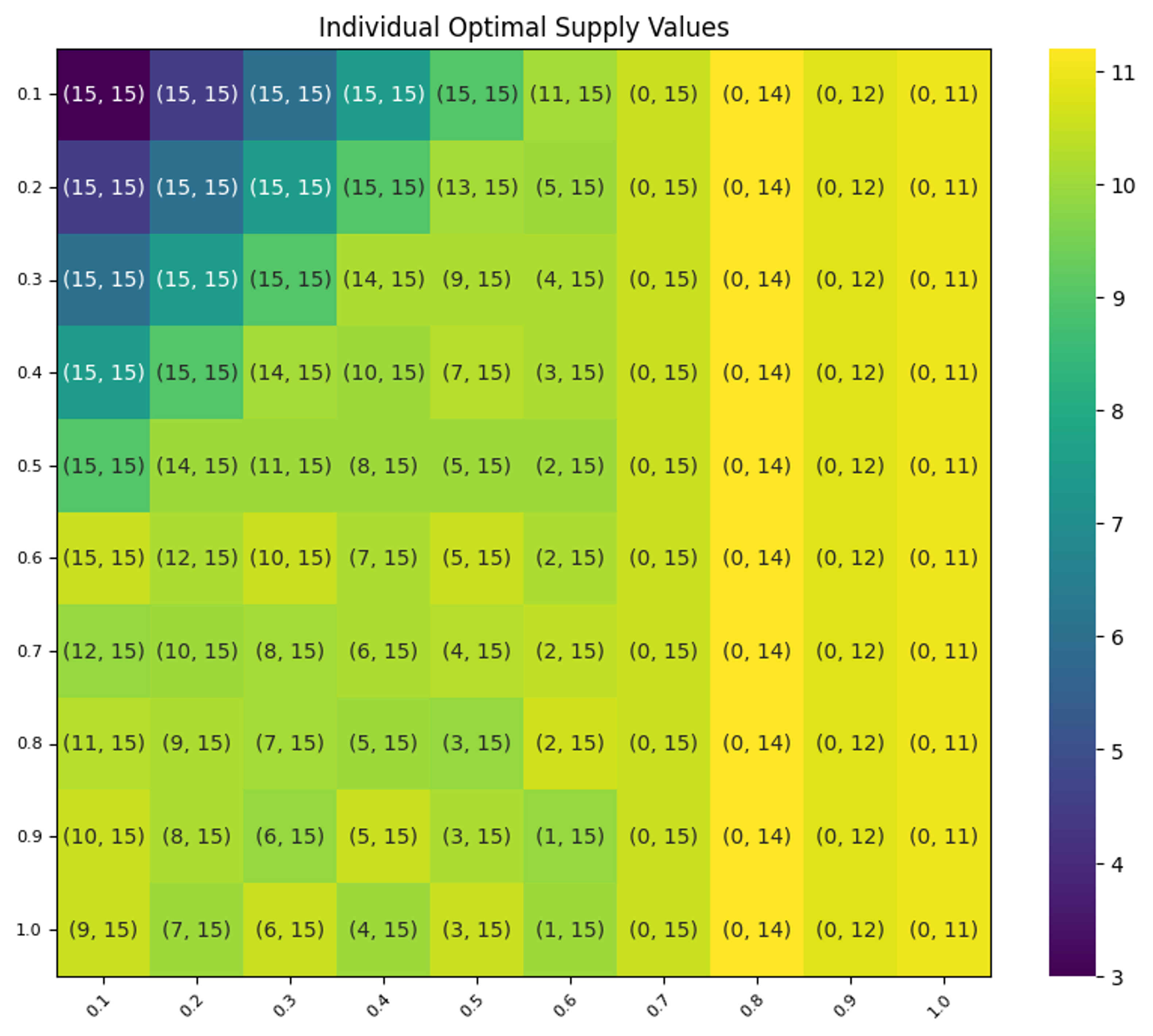}
    \caption{\textbf{Optimal Supply Quantities}: This heatmap illustrates the optimal order quantities for Demand Scenario 1, with the reliability of supplier 2 depicted along the $x$-axis and the reliability of supplier 1 represented along the $y$-axis. 
}
    \label{fig:individual1}
\end{figure}

The heatmap shown in Fig.~\ref{fig:individual1} depicts the optimal order quantities from each supplier, presented as a tuple $(a, b) = (\text{supplier }1, \text{supplier }2)$. Analysis of the heatmap reveals a discernible pattern: since the cost of ordering from supplier 2 ($c_2 = 0.80$) is lower than from supplier 1 ($c_1 = 0.95$), the model tends to allocate a larger portion of the order to supplier 2. This strategic allocation is aimed at maximizing profits, as illustrated by the heatmap.

The observed results align intuitively with the demand distribution shown in Fig. \ref{fig:demand_qgan_1}, which features peaks in the range of $10$ to $13$. Consequently, it is logical to order inventory within this range. Our model, under the assumption of perfectly reliable suppliers (i.e $R_i = 1.0$), recommends a final inventory value of $11.0$. This recommendation is designed to optimize inventory levels for the forthcoming sales season, ensuring sufficient stock to meet the anticipated demand and thereby maximizing profitability.

\subsection{Demand Scenario 2}
\begin{figure}[h!]
    \centering
    \includegraphics[width = \linewidth]{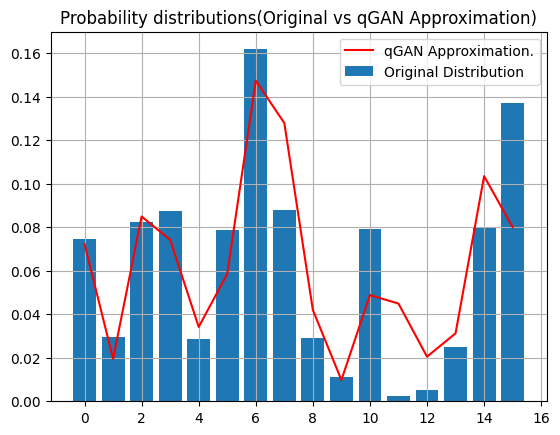}
    \caption{\textbf{Demand Scenario 2}: Comparison of the qGAN learned distribution with the randomly generated demand distribution.}
    \label{fig:demand_qgan_2}
\end{figure}

Utilizing a distinct seed value generates a varied demand scenario, which is characterized by peaks around $6$ and $15$, as depicted in Fig. \ref{fig:demand_qgan_2}. By employing qGANs, we effectively learn this random demand distribution, allowing us to capture its underlying characteristics with high efficiency. 

The optimization procedure employs the COBYLA algorithm, with entropy loss serving as the primary objective function. The results, depicted in Fig.~\ref{fig:demand_qgan_2}, are obtained from qGAN training conducted over 200 epochs. To improve training robustness against the noise inherent in quantum hardware, the optimizer's learning rate has been carefully adjusted to \(10^{-3}\). This fine-tuning ensures effective training dynamics throughout the 200 optimization epochs.

\begin{figure}[h!]
    \centering
    \includegraphics[width=\linewidth]{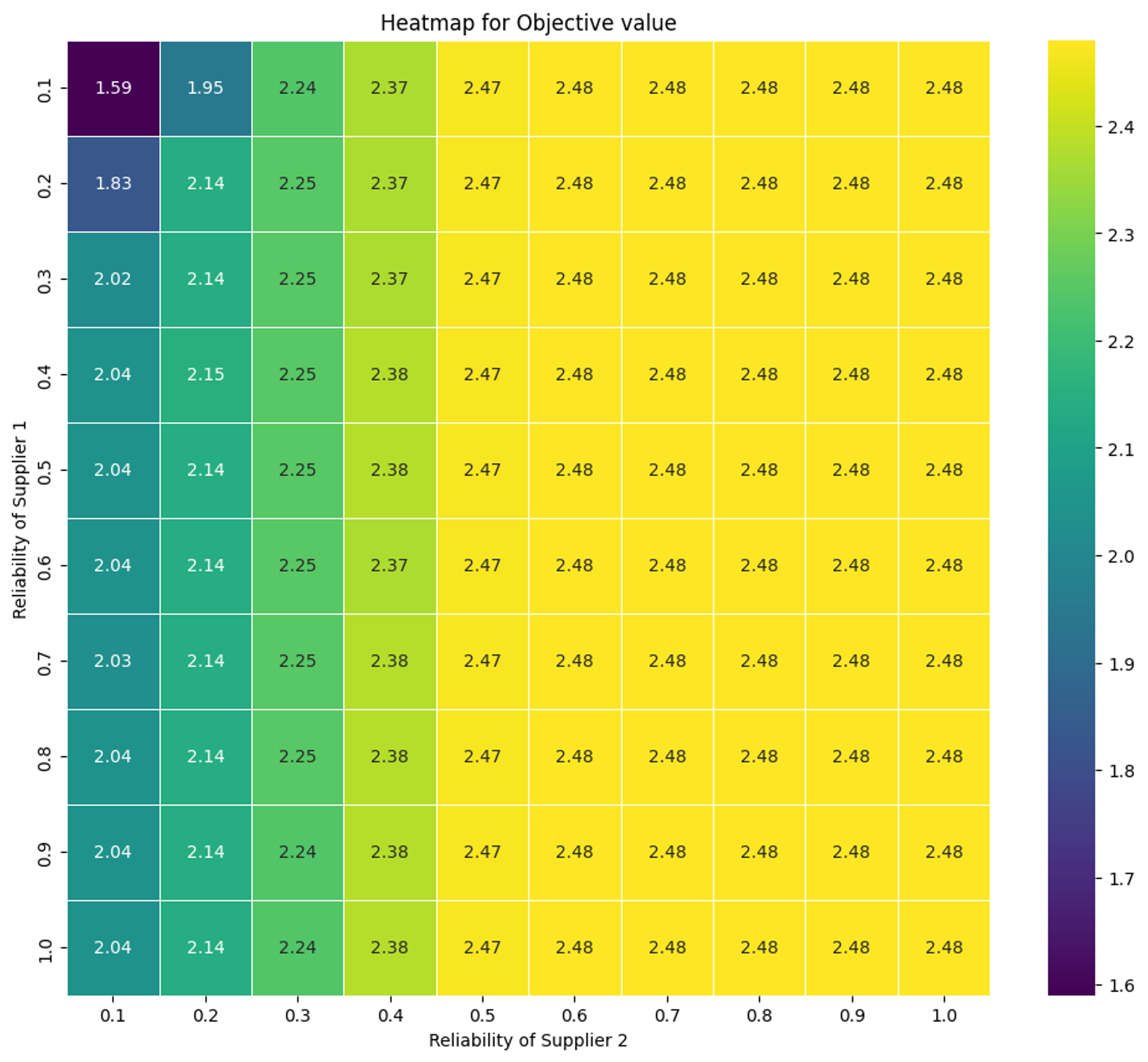}
    \caption{\textbf{Objective Value}: The realized profit for Demand Scenario 2 is illustrated in the heatmap. The \((x, y)\)-axis denotes the reliability of suppliers, while each cell displays the realized profit associated with the corresponding reliability levels of each supplier.
}
    \label{fig:objective2}
\end{figure}

Our experiments entailed varying the reliability factor \( R_i \) for both suppliers, with the results presented in a heatmap format. This heatmap illustrates the objective value, which represents profit, across different reliability factors for each supplier (see Fig.~\ref{fig:objective2}).

\begin{figure}[h!]
    \centering
    \includegraphics[width=\linewidth]{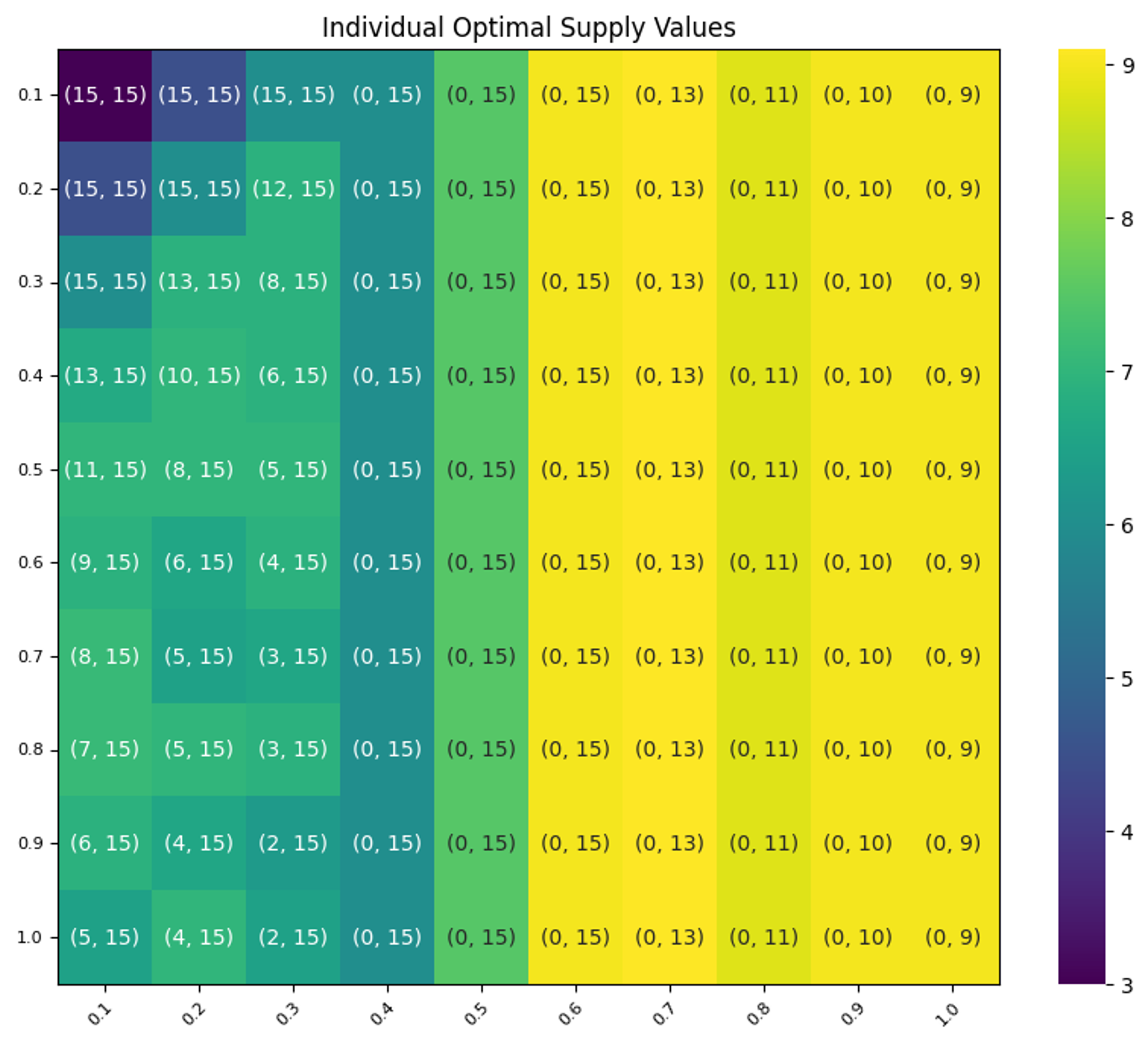}
    \caption{\textbf{Optimal Supply Quantities}: This heatmap illustrates the optimal order quantities for Demand Scenario 2, which favor supplier $2$ due to its lower procurement cost, thereby maximizing profits.
}
    \label{fig:individual2}
\end{figure}

The heatmap presented in Fig.~\ref{fig:individual2} provides insights into the optimal order quantities from each supplier, represented as a tuple \((a, b) = (\text{supplier } 1, \text{supplier } 2)\). A clear pattern emerges from the analysis: due to the lower cost of ordering from supplier 2 (\(c_2 = 0.80\)) compared to supplier 1 (\(c_1 = 0.95\)), the model tends to allocate a larger portion of the order to supplier 2. This strategic allocation is crucial for maximizing profits, as illustrated by the heatmap. This pattern is consistent with the findings for \textbf{Demand Scenario 1}.

The observed outcomes align with the demand distribution depicted in Fig.~\ref{fig:demand_qgan_2}, which shows prominent peaks at values 6 and 15 with minor peaks interspersed. Accordingly, it is logical to place orders within this range, leading our model to suggest a final inventory value of $9.0$ when suppliers exhibit perfect reliability. This strategic decision optimizes inventory levels, ensuring adequate stock availability to meet projected demand for the upcoming sales season.

Furthermore, the results obtained closely correspond with those derived from classical Monte Carlo sampling, demonstrating the robustness and efficacy of Quantum Amplitude Estimation. This consistency highlights the reliability and accuracy of the quantum method in approximating the desired outcomes with high fidelity, providing further validation of its effectiveness in addressing stochastic optimization challenges.

\subsection{Real World Use Case}

In this section, we report results obtained from solving the problem and data provided IBM Manufacturing Solutions Singapore. The goal is to determine the quantity of parts to source from each supplier given their respective reliability.

\begin{figure}[h!]
    \centering
    \includegraphics[width=\linewidth]{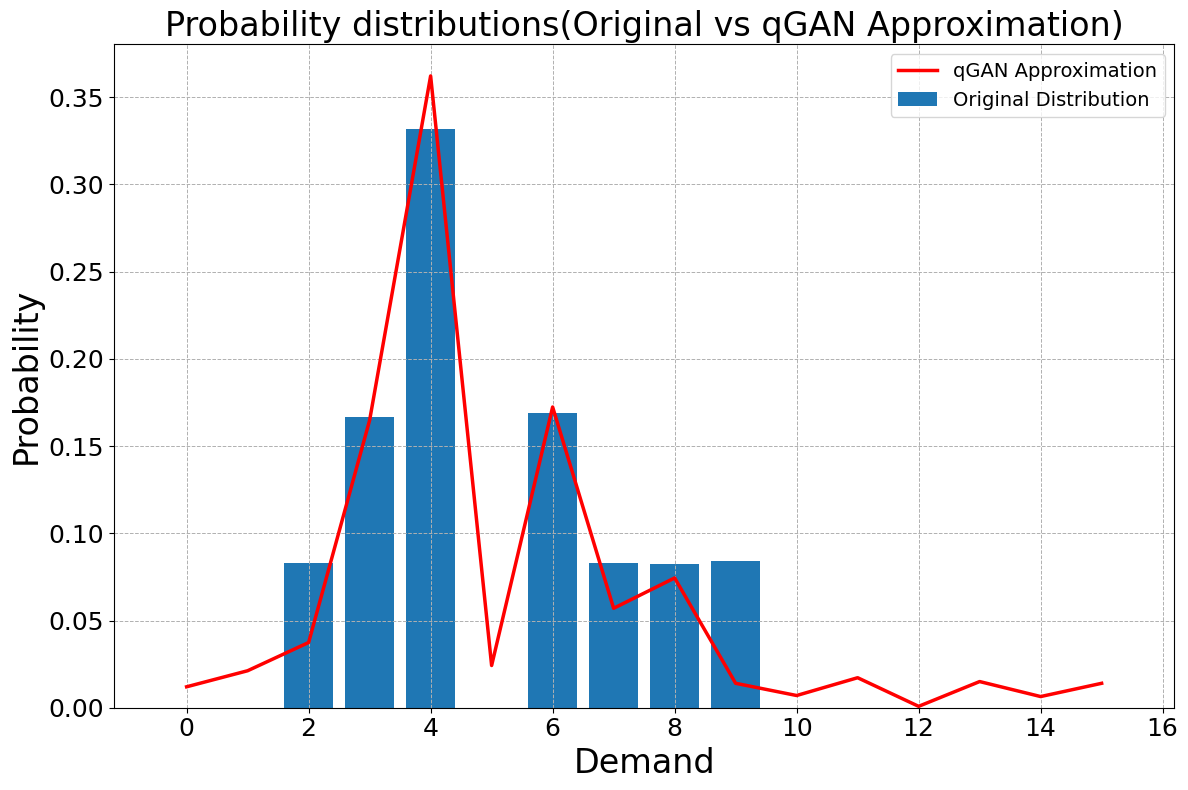}
    \caption{\textbf{Demand Distribution}: Comparison of the qGAN
learned distribution with the IBM Manufacturing Singapore demand distribution.}
    \label{fig:ibm_data}
\end{figure}
Leveraging this data, we trained a qGAN to optimize the quantum circuit parameters required to accurately represent the demand distribution within a quantum framework. See Fig.~\ref{fig:ibm_data}.

The cost and price parameters were provided by IBM, ensuring that the model reflects realistic economic conditions. Meanwhile, other critical parameters, such as supplier reliability, shortage and salvage costs, were inferred from long-term historical data, which allowed us the flexibility to make well-founded assumptions.

To evaluate the effectiveness of this approach, we employed the QAE algorithm. This algorithm enabled us to sample the demand distribution efficiently, compute the corresponding profit, and identify the scenario that maximizes this profit under varying conditions. The quantum circuit used in our study comprises several key components: a probability loading circuit that encodes the demand distribution, a function loading circuit that models the profit function, and a comparator operator \cite{compare} that facilitates the comparison between demand and supply levels.

We experimented with error-mitigated quantum backends. These tests provided insights into the model's performance across different reliability scenarios, which we conceptualized as varying the reliability's variance around the mean to simulate real-world uncertainties.

\begin{figure}[h!]
    \centering
    \includegraphics[width=\linewidth]{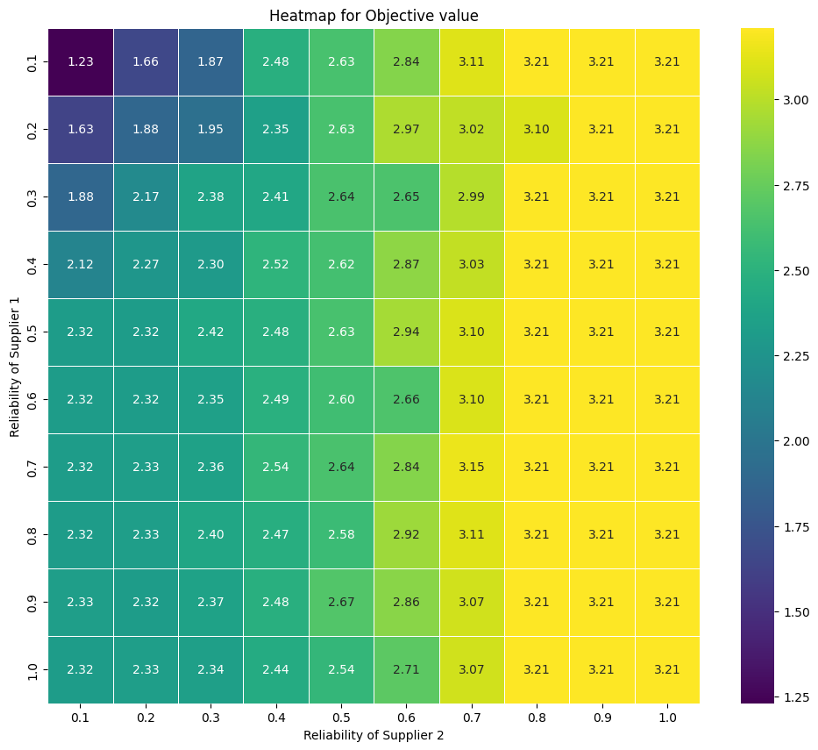}
    \caption{\textbf{Objective Value}: This heatmap represents the objective value of the profit function across different reliability levels for both suppliers using QMC. The \(x\)-axis corresponds to the reliability of the second supplier, while the \(y\)-axis corresponds to the reliability of the first supplier.
}
    \label{fig:ibm_obj}
\end{figure}

\begin{figure}[h!]
    \centering
    \includegraphics[width=\linewidth]{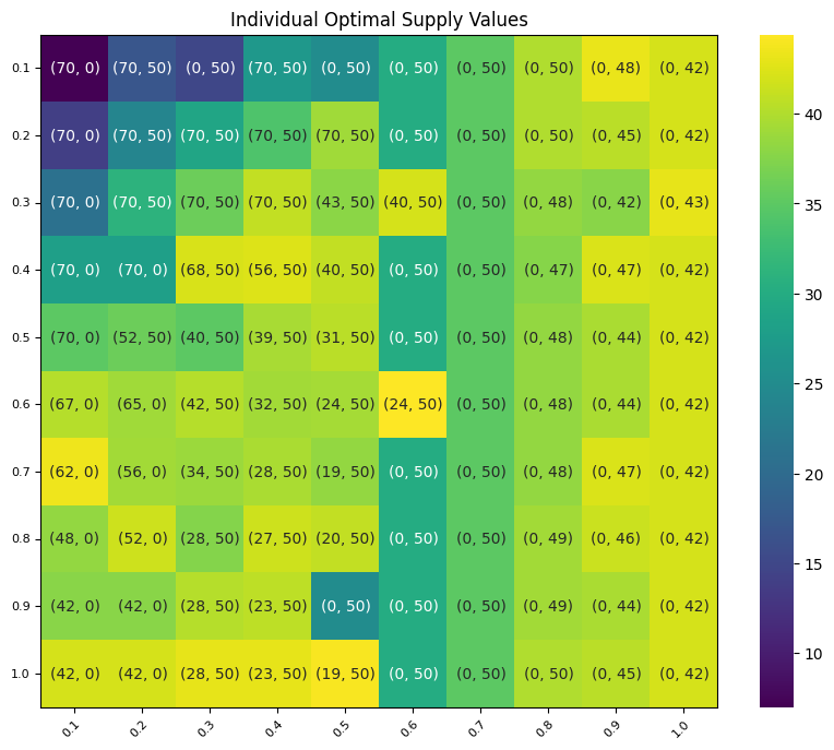}
    \caption{\textbf{Optimal Supply Quantities}: This heat map shows us how much units of the part to order from the suppliers, based on their reliability using QMC. Heatmap showing the model's preference shift from supplier 1 to supplier 2 as supplier 2's reliability increases, despite the latter’s lower cost. The \(x\)-axis corresponds to the reliability of the second supplier, while the \(y\)-axis corresponds to the reliability of the first supplier.}
    \label{fig:ibm_supply}
\end{figure}

The results are presented in Figs.~\ref{fig:ibm_obj} and ~\ref{fig:ibm_supply} which show respectively the objective values and the solutions obtained in terms of supply quantities by our approach. These results reveal a pattern where the model tends to favor supplier 1, despite its higher cost, when the reliability of supplier 2 is relatively low. However, as supplier 2's reliability increases, the model shifts its preference toward supplier 2, driven by its lower acquisition cost.

The optimal amount to be kept at the vendor for the next selling season is $42$ units, as you can see the model tends to reach that value by varying the order quantity between supplier 1 and 2 in low reliability regime, and only ordering from supplier 2 in the high reliability area. 

\begin{figure}[h!]
    \centering
    \includegraphics[width=\linewidth]{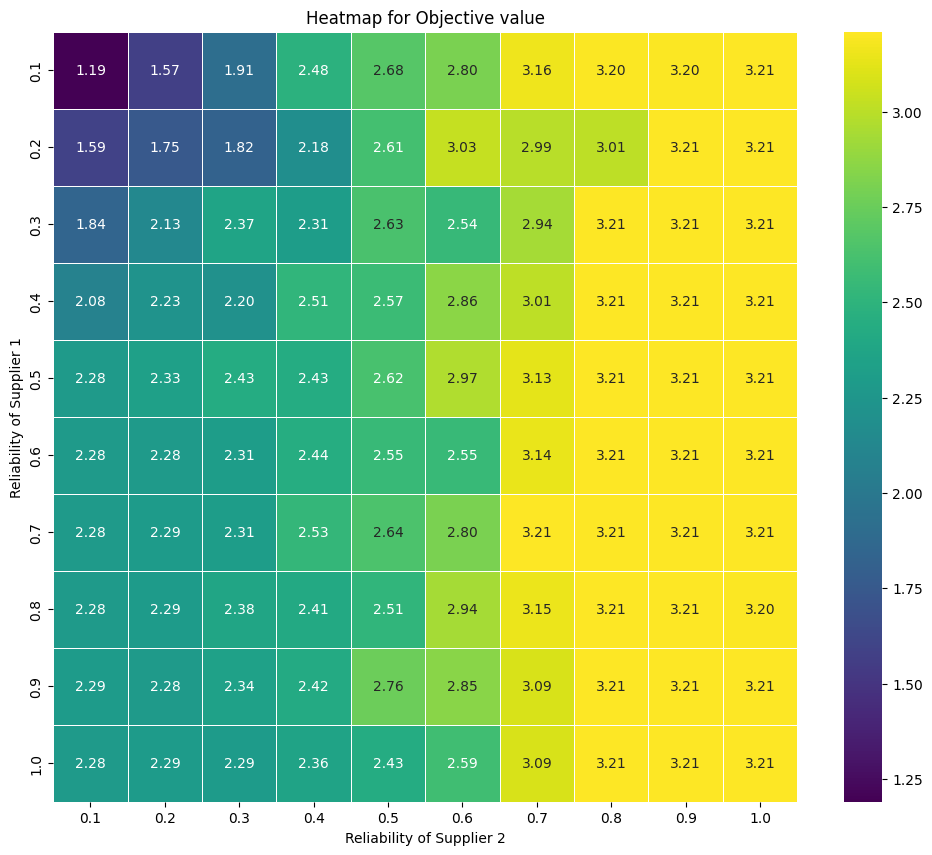}
    \caption{\textbf{Objective Value}: This heatmap represents the objective value of the profit function across different reliability levels for both suppliers using classical SAA method. The \(x\)-axis corresponds to the reliability of the second supplier, while the \(y\)-axis corresponds to the reliability of the first supplier.}
    \label{fig:obj_classical}
\end{figure}

\begin{figure}[h!]
    \centering
    \includegraphics[width=\linewidth]{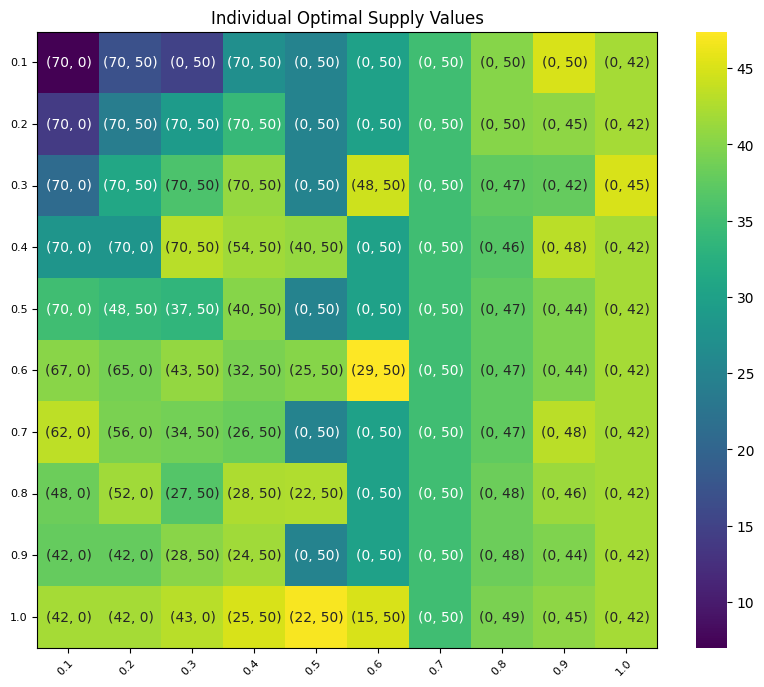}
    \caption{\textbf{Optimal Supply Quantities}: This heat map shows us how much units of the part to order from the suppliers, based on their reliability using classical SAA method. The \(x\)-axis corresponds to the reliability of the second supplier, while the \(y\)-axis corresponds to the reliability of the first supplier.}
    \label{fig:supply_classical}
\end{figure}

Our results align with those obtained using the classical optimization method, Sample Average Approximation (SAA), as shown in Figs.~\ref{fig:obj_classical} and \ref{fig:supply_classical}. The consistency between the quantum and classical approaches underscores the reliability of our method in capturing optimal solutions from real-world data, further validating its practical applicability.

From the heatmaps, we observe that the results obtained from the QAE method closely align with those derived from the SAA method. While a few isolated cells exhibit discrepancies, the overall consistency between the two methods underscores the robustness and reliability of the QAE approach. This agreement suggests that QAE can serve as a viable alternative to classical methods like SAA, particularly in scenarios where quantum algorithms are advantageous. The minor variations observed may warrant further investigation to understand the underlying causes and to refine the quantum algorithm for even greater accuracy.

\section{\label{sec:level8}Conclusion and Future Outlook}

This study addresses the classical single-period inventory problem in the context of multiple unreliable suppliers using QMC methods. The primary focus is on exploring the complexities introduced by engaging with multiple global suppliers, which may result in sourcing risks and potential delays in demand fulfillment. The model enables the newsvendor to strategically select a subset of suppliers from a pool of candidates and to allocate product requirements among these chosen suppliers. Additionally, the framework incorporates a fixed plus linear cost structure for each supplier, where the fixed cost includes transportation expenses and the initial investment required to establish a business relationship with a new supplier.

The decision-making process for supplier selection in this context relies on a detailed assessment of variable order costs, fixed ordering costs, and the anticipated reliability of suppliers. Sensitivity analysis indicates that unit ordering cost is the most influential factor in supplier selection, overshadowing considerations of reliability. As supplier reliability improves, there is a noticeable shift toward reducing the number of suppliers, favoring a more consolidated approach. This shift is driven by the advantage of ordering larger quantities from fewer suppliers, which is supported by lower procurement costs. Conversely, higher shortage and salvage costs encourage a multi-sourcing strategy, highlighting the complex interaction between cost factors and supplier reliability in shaping the newsvendor's decisions.

Recent advancements have led to the development of advanced techniques for state preparation and loading, enabling the direct encoding of all components of a real-valued data vector into the amplitude of a quantum state. Unlike previous methods, which could only load the absolute values of these components, these new techniques offer potential benefits for applications such as financial pricing models and inventory management data loading \cite{newdata}. These methods provide an alternative to quantum Generative Adversarial Networks (qGAN) for more efficient loading of probability states.

Despite these advancements, several unresolved issues remain that warrant further investigation. One such issue involves identifying optimal quantum generator and discriminator structures, as well as refining associated training methodologies. Certain structural configurations may exhibit superior performance for specific tasks, potentially leading to optimal outcomes. Addressing these questions will advance quantum data loading techniques and their applications across various domains.

Additionally, the use of Quantum Amplitude Estimation (QAE) for solving the Newsvendor problem requires further scrutiny. Given its status as a quantum algorithm that may not consistently offer a quantum advantage, particularly in the pre-fault tolerance era, it is essential to consider that implementing QAE could involve significant computational expenses due to the number of quantum gates required. Moreover, there are scenarios where QAE may not outperform classical algorithms in practical applications. This highlights the need for a thorough comparative analysis of quantum and classical approaches to effectively address inventory management challenges.

\section{Acknowledgement}

This research is supported by the National Research Foundation, Singapore under its Quantum Engineering Programme 2.0 (NRF2021-QEP2-02-P01). We thank IBM Manufacturing Solutions Singapore for providing the problem statement and data.

\newpage

\newpage
\bibliography{reference}


\appendix
\newpage

\section{Newsvendor Formulation}\label{form}

The newsvendor model is a widely used mathematical framework in operations management aimed at establishing the most effective inventory levels. It operates under the premise of fixed prices and unpredictable demand for a perishable item. In this model, if the inventory level is set at $Q$, any excess demand beyond $Q$ results in missed sales opportunities, while any shortfall in demand below $Q$ is sold at a unit salvage price.

The newsvendor problem is a common occurrence in various business sectors and industries. It enables decision-makers to determine the optimal order quantity at present, despite facing uncertainty regarding future demand.

All the newsvendor models have a common mathematical structure with the following elements:
 \begin{itemize}
     \item A Decision Variable ($Q$) : The newsvendor problem is to find the value $Q$ that leads to an optimal decision. This value of $Q$ is denoted by $Q^{\star}$

     \item Uncertain Demand(D) : Demand is a random variable defined by the demand distribution and estimates of the parameters of the demand distribution.

     \item Unit Overage Cost ($c_o$) : This is the cost of buying on unit more  than the demand during a selling season. Also $c_o = c-s$, where $c$ is  the unit cost and $s$ is the unit salvage value (the value  of an asset at the end of its life)

     \item Unit Underage Cost ($c_u$) :  This is the cost  of buying one unit less than the demand during a selling season. Also, $c_u = p-c$, where $p$ is the unit price.
 \end{itemize}

In an optimization problem, the goal is to minimize the loss function or the cost function, which serves as the objective function. In the context of the newsvendor model, the cost function is minimized to find the optimal quantity $Q^{\star}$. The expected cost $\mathbb{E}[\text{Cost}(Q)]$ is a convex function, indicating that its minimization occurs. This assumption is made under the condition that $D$ is a continuous random variable with a density function $f(D)$ and a cumulative distribution function $F(D)$.

The cost is given by 

\[
    \text{Cost}(Q,D) = 
\begin{cases}
    c_o(Q-D),& \text{if } D\leq Q\\
    c_u(D-Q),              & \text{otherwise}
\end{cases}
\]

and the expected cost is given by 

\begin{align*}
    \mathbb{E}[\text{Cost}(Q)]  &= \int_{D=0}^{\infty} \text{Cost}(Q,D) f(D) dD \\
    & = c_o \int_{0}^{Q} (Q-D)f(D) dD\\
    & \ \ + c_u \int_{Q}^{\infty} (D-Q)f(D) dD  \\
     & = c_o Q \int_{0}^{Q} f(D) dD - c_o \int_{0}^{Q} Df(D) dD \\
     & \ \ + c_u \int_{Q}^{\infty} Df(D) dD -  c_u Q \int_{Q}^{\infty} f(D) dD \\
     & = c_o QF(Q) - c_o T(Q) + c_u (\tau - T(Q)) \\
     \\
     & -  c_u Q(1-F(Q)) \\
     \\
     & = (c_u + c_o) (QF(Q) - T(Q)) + c_u(\tau - Q)
\end{align*}

Here $F(Q)$ is the demand distribution function evaluated at $Q$ and

$$ \int_{D=Q}^{\infty} D f(D) dD = \tau - T(Q)$$

We use Leibniz's rule stated as:

If $f$ is continuous on $[a,b]$ and if $u(x)$ and $v(x)$ are differentiable functions of $x$ whose values lie in $[a.b]$ then

$$ \frac{d}{dx} \int_{u(x)}^{v(x)} f(t) dt = f(v(x)) \frac{dv}{dx} -  f(u(x)) \frac{du}{dx}$$

Taking the first derivative of $\mathbb{E}[\text{Cost}(Q)]$ with respect to $Q$, we have

\begin{align*}
    \frac{d \mathbb{E}[Cost](Q)}{dQ}  & = (c_u + c_o)(QF'(Q) + F(Q) - T'(Q)) - c_u \\
    & =  (c_u + c_o)(Qf(Q) + F(Q) - Qf(Q)) - c_u  \\
    & = (c_u + c_o)F(Q) - c_u
\end{align*}

Equating this derivative to zero, we have 

$$ F(Q) = \frac{c_u}{c_u + c_o}$$

where $\alpha = \frac{c_u}{c_u + c_o}$ is called the critical fractile or critical ratio.
The critical fractile is what we get when the cost is at its minimum and $Q^{\star} = F^{-1}\big( \frac{c_u}{c_u + c_o}\big)$

\newpage
\section{Monte Carlo Methods}\label{montecarlo}

Monte Carlo methods are statistical techniques employed to approximate solutions for tasks like computing expected values of functions or integrating functions that resist analytical integration \cite{integration}. These methods hinge on our capacity to randomly sample a variable based on its probability distribution.

The principles of Monte Carlo methods are based on the strong law of large numbers defined below:

\textbf{The Strong Law of Large Numbers} : If $X_1, X_2, .... , X_n$ are \textit{independent} and \textit{identically distributed} random variables with $\mathbb{E}[X_k] = \mu$ for $k=1,2,....$  ,then

$$ P \Bigg(\lim_{n \rightarrow \infty} \frac{\sum_{k=1}^{n}X_k}{n} \Bigg) = 1$$
The law of large numbers says that the sample mean approaches the theoretical mean as the number of identically distributed, randomly generated variable increases.

The expectation of a continuous random variable $X$ with probability density function $f(x)$ is the number

$$ \mu = \mathbb{E}[x] = \int_{-\infty}^{+\infty} x f(x) dx$$

provided the integral

$$ \int_{-\infty}^{+\infty} |x| f(x) dx$$

is finite.

Suppose we have an integral 

$$ I = \int_{a}^{b} g(x) f(x) dx$$

that needs to be evaluated. The Monte Carlo method is to take a random sample $X_1, X_2, ...,X_n$ from this distribution, and then form the mean

$$ g_n = \frac{1}{n} \sum_{k=1}^{n} g(X_k)$$

From the strong law of large numbers, we have that 

$$ \frac{1}{n} \sum_{k=1}^{n} g(X_k) \rightarrow \int_{a}^{b} g(x) f(x) dx$$

with probability $1$. For a better evaluation of the integral, we need to make $n$ as large as possible.\\

\section{{Payoff Function}}\label{payoff}

We create the operator $F$ that maps $\sum_i \sqrt{p_i} |i\rangle_n |0\rangle$ to

\begin{equation*}
    \sum_{i=0}^{2^n-1} \sqrt{p_i}|i\rangle_n \bigg[ \cos \Big(c \tilde{f}(i) + \frac{\pi}{4}\Big)|0\rangle + \sin\Big( c \tilde{f}(i) + \frac{\pi}{4}\Big) |1\rangle \bigg]
\end{equation*}

where $\tilde{f}(i)$ is just the scaled version of $f(i)$ given by:

\begin{equation}{\label{eq:d1}}
    \tilde{f}(i) = 2 \frac{f(i) - f_{min}}{f_{max} - f_{min}}-1
\end{equation}

with $f_{min} = \text{min} f(i)$ and $f_{max} = \text{max}f(i)$ and $c \in [0,1]$ is an additional scaling parameter. With these definitions, the probability of finding the ancilla qubit in state $|1\rangle$ namely

\begin{equation}
    P = \sum_{i=0}^{2^n -1} p_i \sin^2 \bigg( c \tilde{f}(i) + \frac{\pi}{4} \bigg)
\end{equation}

is well approximated by

\begin{equation}
    P \approx \sum_{i=0}^{2^n-1} p_i \bigg( c \tilde{f}(i)+ \frac{1}{2} \bigg) = c \frac{2 \mathbb{E}[f(X)] - f_{min}}{f_{max}- f_{min}} - c + \frac{1}{2}
\end{equation}

To obtain this result we made use of the approximation

\begin{equation}
    \sin^2 \bigg( c \tilde{f}(i) + \frac{\pi}{4} \bigg) = c\tilde{f}(i) + \frac{1}{2} + \mathcal{O}(c^3 \tilde{f}^3 (i))
\end{equation}

which is valid for small values of $c \tilde{f}(i)$. With this first-order approximation, the convergence rate of QAE is $\mathcal{O}(M^{-2/3})$ when $c$ is properly chosen which is already faster than the classical Monte Carlo methods \cite{risk}. We can recover the $\mathcal{O}(M^{-1})$ convergence rate of QAE by using higher orders implemented with quantum arithmetic. The resulting circuits, however, will have more gates. This trade-off is discussed in \cite{risk}, and also gives a formula that specifies which value of $c$ to use to minimize the estimation error made when using QAE.

\end{document}